\documentclass[aps,twocolumn,a4paper,pra,superscriptaddress,floatfix]{revtex4}
\usepackage{epsfig}
\usepackage{amsmath}%
\usepackage{amsfonts}%
\usepackage{amssymb}%
\usepackage{graphicx}

\newcommand\bk{\mathbf{k}}

\begin{document}

\title{Casimir-Polder interaction between an atom and a dielectric grating}
\author{Ana M. Contreras-Reyes} 
\affiliation{Instituto de Fisica, UFRJ, CP 68528, Rio de Janeiro, RJ, 21941-972, Brazil}  
\author{Romain Gu\'erout}
\affiliation{Laboratoire Kastler Brossel, case 74, CNRS, ENS, UPMC, Campus Jussieu, F-75252 Paris Cedex 05, France}
\author{Paulo A. Maia Neto}
\affiliation{Instituto de Fisica, UFRJ, CP 68528, Rio de Janeiro, RJ, 21941-972, Brazil}  
\author{Diego A. R. Dalvit}
\affiliation{Theoretical Division, Los Alamos National Laboratory, Los Alamos, NM 87545, USA}
 \author{Astrid Lambrecht}
 \author{Serge Reynaud}
\affiliation{Laboratoire Kastler Brossel, case 74, CNRS, ENS, UPMC, Campus Jussieu, F-75252 Paris Cedex 05, France}
\date{\today}
\begin{abstract} We develop the scattering approach to calculate the exact dispersive Casimir-Polder potential between a ground-state atom and a rectangular grating. 
Our formalism allows, in principle, for arbitrary values of the grating amplitude and period, and of the atom-grating distance. We compute numerically the potential 
for a Rb atom on top of a Si grating and compare the results with the
potential for a flat surface taken at the local atom-surface distance 
(proximity force approximation).  Except for very short separation distances, the potential  is nearly sinusoidal along the direction transverse to the 
grooves. 
\end{abstract}
 \maketitle

\section{Introduction}

Vacuum field fluctuations are modified close to material surfaces, resulting in the usually attractive Casimir-Polder force \cite{Casimir48} on a nearby 
 ground-state atom.  This effect has been measured by a number of  experimental 
 techniques,  including deflection of atomic beams \cite{Sukenik93}, 
 classical \cite{Landragin96,Bender10} and quantum reflection by the attractive Casimir-Polder
 potential \cite{Shimizu01,Druzhinina03,Pasquini04}, and dipole oscillations of a 
 Bose-Einstein condensate (BEC)  close to a dielectric surface \cite{Harber05}. 
 
More recently, experiments involving non-trivial geometries paved the way to new applications. 
  The measured reflection probability of a BEC from 
  a Si surface with a square array of closely spaced thin pillars
increased by a factor of nearly four as compared to a flat surface  \cite{Pasquini06}. 
A  Si surface with an array of wall-like  parallel ridges was shown to work as a reflection diffraction grating for atoms incident along a nearly 
 grazing direction \cite{Oberst05}.
Several diffraction orders in the quantum reflection from a microstructured grating consisting of Cr strips on a flat quartz substrate were measured~\cite{Zhao08}.
Alternatively, nano-fabricated transmission atom gratings allow for a direct measurement of the dispersive atom-surface potential in the 
non-retarded van der Waals regime \cite{Perreault05}.

Remarkable experimental progress has also been achieved in the closely related field
of Casimir interactions between material surfaces \cite{Capasso07}. 
Recent experiments have revealed interesting 
geometry effects in  the normal force between a Si 
rectangular (lamellar) grating and a metallic spherical surface \cite{Chan08, Bao10} 
and in the lateral force between two metallic gratings \cite{Chiu09}. 

This ensemble of new experiments clearly motivates 
the theoretical analysis of how  geometry molds the quantum field fluctuations giving  rise to the Casimir-Polder and Casimir interactions. 
In this paper, we
develop a non-perturbative theory for the Casimir-Polder  interaction between a ground state atom and a dielectric grating. Our results are valid, in principle, 
for arbitrary values of the grating amplitude. Preliminary results were already applied  in the analysis of quantum vortex generation in a BEC induced 
 by the Casimir-Polder potential of a rotating grating~\cite{Impens10}.  Here we present a detailed derivation of the Casimir-Poder potential and 
 a variety of numerical examples that illustrate  its main characteristics.  

 Calculations beyond the simple planar geometry are extremely involved because 
different frequency and length scales contribute to the interaction. 
Since dispersive forces are not additive, it is not possible to build up the atom-surface potential
from the more elementary atom-atom interaction. In the pairwise summation (PWS) approach, the non-additivity is corrected by
a `calibration' provided by the planar case at a given separation distance range \cite{Bezerra00}. Since the non-additivity correction is
geometry and distance dependent, PWS results are barely more accurate \cite{Dalvit08} than the
 results obtained by taking the proximity force approximation (PFA) \cite{Derjaguin57}, 
in which the potential is simply approximated by the planar case taken at the local atom-surface distance. 

Only recently new theoretical  tools have been developed for
analyzing non-trivial geometries~\cite{review} beyond the PFA and PWS approaches. In the 
scattering approach \cite{Lambrecht06,Emig07}, 
the interaction potential is written in terms of  reflection operators describing non-specular scattering (diffraction)
by   non-planar  surfaces.  This allows for a  description of the Casimir  effect that captures  its full geometry dependence. 
In the specific case of the Casimir-Polder interaction between a ground-state atom and a material surface, the exact potential can be calculated 
for arbitrary distances, taking due account of finite response times of both atom and material medium, provided the surface reflection operator is known~\cite{Messina09}. 

By computing the reflection operator of the grating as a  perturbation of the planar symmetry, 
Refs.~\cite{Messina09, Dalvit08A} derived the Casimir-Polder potential  to first order in the grating amplitude. 
The resulting expressions  are valid only when the grating amplitude is the smallest length scale in the problem. 
However, in order to enhance the non-trivial geometry effects associated with the departure from the planar symmetry, it is of course  interesting to have
large amplitudes. This was indeed the case in the atom-surface  \cite{Pasquini06,Oberst05}
and surface-surface \cite{Chan08,Chiu09} experiments carried out so far.  In this paper, we
compute the  exact 
 reflection operator  by applying the differential theory  of diffraction gratings 
 \cite{Neviere98}, and then deriving the exact Casimir-Polder potential from the scattering approach  \cite{Messina09}.
Our approach is
 similar to the theory developed in Refs.~\cite{Lambrecht08, Chiu10, Lambrecht10}  for treating surface-surface interactions.

Non-perturbative results were previously obtained for a 
 toy model describing the Casimir interaction of a
scalar field satisfying ideal Dirichlet boundary conditions 
with a small sphere above  a corrugated 
surface~\cite{Dobrich08}. In this paper, we develop a full electromagnetic theory that takes into account
the electromagnetic responses of real atoms and material surfaces (we consider a Rb atom  above a Si grating in the numerical examples). 
This allows us to cover the entire range of separation distances, from the unretarded short-distance van der Waals  regime to the Casimir-Polder large-distance asymptotic limit. 

The paper is organized as follows. In Sec. II, we derive the formal results for the Casimir energy which are applied in the numerical implementations discussed in 
Sec. III. Sec. IV presents our concluding remarks. 

\section{Scattering approach to the atom-grating interaction}

We consider a spherically symmetric ground-state atom located at $(x_A,y_A,z_A)$ above a non-planar surface (see Fig.~\ref{system}), 
corresponding to a profile function $h(x,y)$ giving the local surface height with respect to a reference plane at $z=0.$  
 It is convenient to develop the scattering formula in the plane-wave basis $|\bk, \pm, p\rangle$  \cite{Lambrecht06}, where $\bk$ is the two-dimensional wave-vector component parallel to the 
 $xy$ plane, $+(-)$ represents upward (downward) propagation direction and $p$ stands for polarization. 
 We assume
that the atom-surface separation distance $z_A-h(x_A,y_A)$ is much larger than the atomic dimensions, 
allowing us to approximate the potential  to first order in the atomic  polarizability $\alpha (\omega)$ or, equivalently, 
to first order in the atomic
 reflection operator \cite{Messina09}
  \begin{widetext}
 \begin{equation} \label{Ratomic}
 \langle \mathbf {k}, -,p | \mathcal R_A (i \xi) | \mathbf {k'},+, p' \rangle = - \frac{ \xi^2}{2 \kappa} \frac{\alpha (i \xi)} {\epsilon_o c^2}
  e^{i (\mathbf {k} -\mathbf {k'}) \cdot \mathbf{r}_A  }e^{- (\kappa + \kappa') z_A}
\hat  {\boldsymbol {\epsilon}}_{p}^- (\mathbf {k}, i \xi) \cdot \hat  {\boldsymbol {\epsilon}}_{p'}^+ (\mathbf {k'}, i \xi'),
\end{equation}
 with 
 $ \mathbf{r}_A=(x_A,y_A)$ and
 $\kappa=\sqrt{k^2+\xi^2/c^2}$ representing the wave-vector $z$-component associated with the imaginary frequency $\xi$ ($\kappa'$ is defined in terms of $\bk'$ in the same way). $ \hat  {\boldsymbol {\epsilon}}_{p}^{\pm}(\mathbf {k}, i \xi)$
 are  unit vectors corresponding to a given polarization basis (to be chosen later in this section).
 
\begin{figure}[h]
\begin{center}
\bigskip\bigskip
\includegraphics[scale=.33]{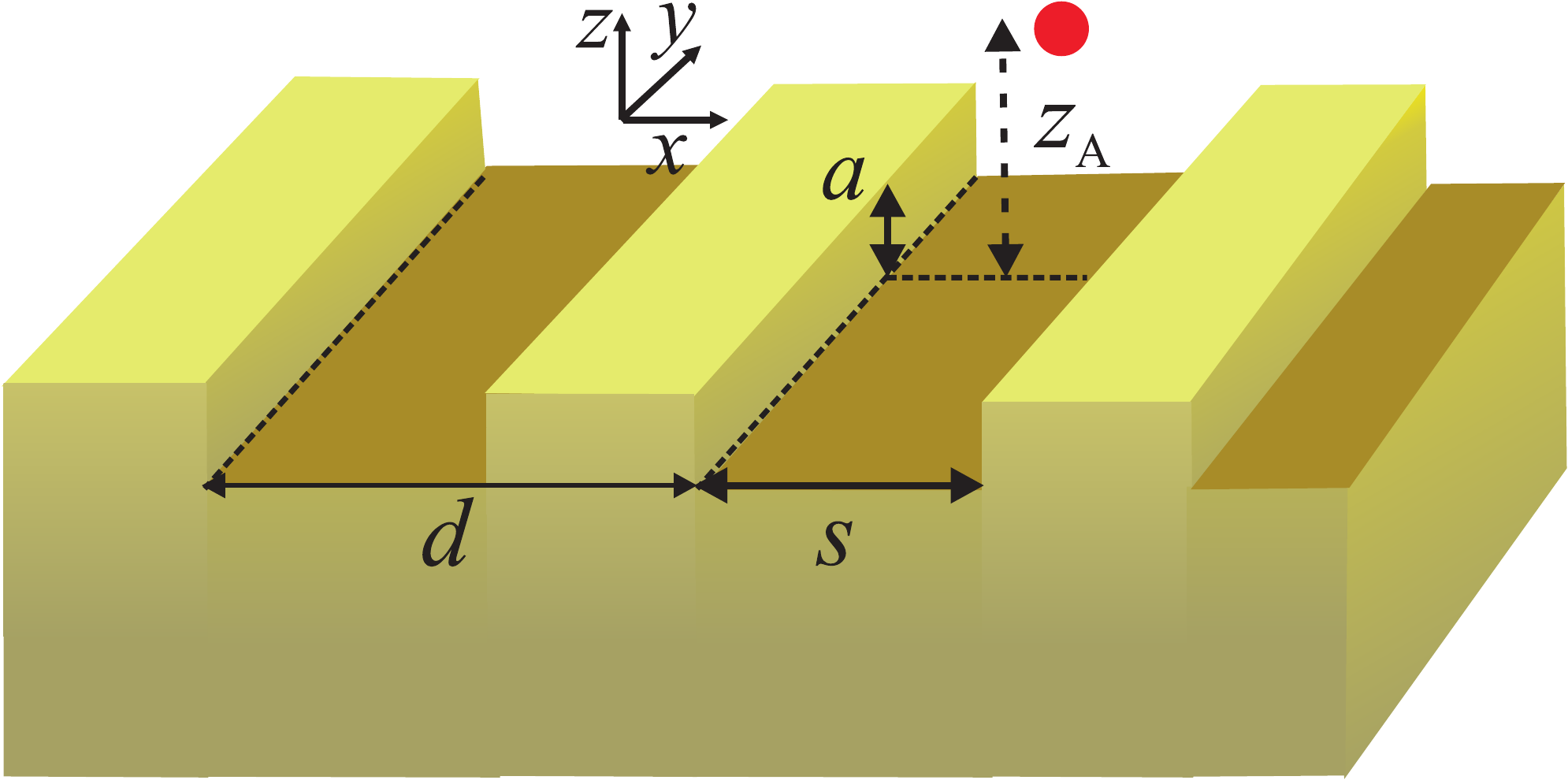}
\end{center}
\caption{\label{system} \footnotesize{Atom on top of a rectangular grating.}} 
\end{figure}

 The zero-temperature Casimir energy is then obtained by expanding the general scattering formula \cite{Lambrecht06, Emig07} to first order in
 ${\cal R}_A.$
 This corresponds to neglecting all multiple reflections between the  atom and the surface (reflection operator ${\cal R}_S$) but the 
 single round-trip containing one reflection by each one  (see Ref.~ \cite{Messina09} for a detailed derivation).
 The Casimir potential is then written as an integral over the positive imaginary  semi-axis in the complex frequency plane:
 \begin{equation} 
 \label{interaction}
U(x_A,y_A,z_A) = - \hbar \int_0^{\infty} \frac{\rm d \xi}{2 \pi} \int \frac{\rm d^2 \mathbf {k}}{(2 \pi)^2}\int  \frac{\rm d^2 \mathbf {k'}}{(2 \pi)^2}
\sum_{p, p'}\langle \mathbf{k},+, p | \mathcal R_S (i \xi) | \mathbf {k'},-, p' \rangle   
\langle \mathbf{k'},-, p' | \mathcal R_A(i \xi) | \mathbf{k},+, p \rangle. 
\end{equation}
 \end{widetext} 
 
 An exact expression for the Casimir-Polder interaction energy between a
ground-state atom and a generic surface can also be written in terms of an
integral over real frequencies. Once such an integral is computed with an
appropriate choice of integration contour in the complex frequency plane,
one recovers Eq. (\ref{interaction}), which is in terms of purely imaginary frequencies
$\omega=i \xi$. On one hand, the real frequency formalism suffers from
highly oscillatory contributions, but allows for the identification of the
relevant modes (evanescent and propagating) of the electromagnetic field
that contribute to the interaction energy. For example, both modes are known to play a fundamental role
in the Casimir energy in the plane-plane geometry \cite{Henkel2004}. We expect that both kinds of modes also have a key role in the
Casimir-Polder atom-grating interaction (this analysis is beyond the
scope of the present paper). On the other hand, the formalism in terms of purely
imaginary frequencies has a smooth, exponentially decaying frequency
behavior, but it does not allow for the identification of the separate contributions from evanescent and propagating modes. However, Eq. ~(\ref{interaction})
does include the total contribution of all modes.

 In this paper, we calculate the exact surface reflection operator ${\cal R}_S$ for the rectangular (lamellar) grating  shown in Fig.~\ref{system}.  
The periodic profile function $h(x)$  has   period $d,$ amplitude $a$ and corresponds to a  groove width  $s.$ The reference plane $z=0$ is located at 
the bottom of the groove. 
Translational symmetry along the $y$ direction implies that ${\cal R}_S$ does not change $k_y.$ When defining the polarization basis, 
we exploit this symmetry as in waveguide theory:  polarization H is such that $E_y\equiv 0$ with $E_x,$ $E_z,$ $H_x$ and $H_z$ given in terms of $H_y$ by using Maxwell 
equations. For the electric field components at a real frequency $\omega$ propagating in a medium of dielectric constant $\epsilon$, we have
$E^{\rm H}_x=-ic\omega\partial_zH_y/(\epsilon\omega^2-c^2k_y^2)$ and $E^{\rm H}_z=ic\omega\partial_xH_y/(\epsilon\omega^2-c^2k_y^2).$
In vacuum ($\epsilon=1$), the corresponding unit vectors are then given by 
\begin{equation}\label{epsilonH}
\hat {\boldsymbol {\epsilon}}_ {\rm H}^{\pm}(\bk,\omega) = \frac{1}{\sqrt{\omega^2/c^2-k_y^2}}(\mp \sqrt{\omega^2/c^2-k^2}
\,\mathbf{\hat x} + k_x\,\mathbf{\hat z}).
\end{equation}
Polarization E is likewise defined by the condition $H_y\equiv 0,$
with  $E^{\rm E}_x=ic^2k_y\partial_xE_y/(\epsilon\omega^2-c^2k_y^2)$ and 
$E^{\rm E}_z=ic^2k_y\partial_zH_y/(\epsilon\omega^2-c^2k_y^2),$
so that the corresponding unit vectors are given by
\begin{eqnarray}
\nonumber
\hat {\boldsymbol {\epsilon}}_ {\rm E}^{\pm}(\bk,\omega) &=& \frac{c}{\omega}\Biggl[
\frac{k_y}{\sqrt{\omega^2/c^2-k_y^2}}   \left( -  k_x\, \mathbf{\hat x}  \mp  \sqrt{\omega^2/c^2-k^2}\,\mathbf{\hat z}\right)
\\
\label{epsilonE}
 & & +   \sqrt{\omega^2/c^2-k_y^2}\,\mathbf{\hat y}
\Biggr].
\end{eqnarray}

 Since the surface profile is periodic along the $x$ direction, the fields are pseudo-periodic functions of $x$ 
(Bloch's theorem): ${\bf E}(x+d,y,z)= e^{ik_x^{(0)} d} {\bf E}(x,y,z)$ for some $k_x^{(0)}$ in the first Brillouin zone $[-\pi/d,\pi/d].$ 
The wave-vector $x$-component of a given  incident plane wave can always be cast in the form 
\begin{equation}
\label{kxj}
k_x^{(j)} = k_x^{(0)}+ j \,\frac{2\pi}{d}
\end{equation}
for some integer $j$ and $k_x^{(0)}\in [-\pi/d,\pi/d].$ Diffraction by the periodic grating will give rise to new Fourier components modulated by integer multiples of $2\pi/d.$ 
Although these new Fourier components correspond to
different integers $j'$ (with $j'-j$ representing a given diffraction order), they all  correspond to the
 the same $k_x^{(0)}$ in the first Brillouin zone according to  Bloch's theorem. 
 For a given incident plane wave $| (k_x^{(0)}+ j \,\frac{2\pi}{d}) \mathbf{\hat x}+k_y  \mathbf{\hat y},-,p\rangle,$ we add, 
  in the homogenous region above the grating ($z\ge a$),  a superposition of  reflected plane waves $| (k_x^{(0)}+ j' \,\frac{2\pi}{d}) \mathbf{\hat x}+k_y  \mathbf{\hat y},+,p'\rangle$
 (Rayleigh expansion), with  amplitudes given by the matrix elements
 $ \langle j', p' | \mathcal R_S  | j, p \rangle$  of the grating reflection operator.
 In the bulk region $z<0,$ the dielectric constant is also uniform and the field is  written as a simple plane wave expansion in terms of transmission amplitudes. 
 On the other hand, within the inhomogeneous grating region $0\le z\le a,$ we derive the non-trivial $z$ dependence of the field by solving coupled differential equations
  (differential approach) \cite{Neviere98}.  We then solve for the  matrix elements $ \langle j', p' | \mathcal R_S | j, p \rangle$ by matching
  the different field expansions at $z=a$ and $z=0.$  
  
  The properties of the reflection operator ${\cal R}_S$ discussed above allow us to simplify the general expression (\ref{interaction})
  for the Casimir-Polder potential energy.  By plugging the result 
  (\ref{Ratomic})
  for the atomic reflection operator into 
  (\ref{interaction}), we find
 \begin{eqnarray} \label{final}
 && U(x_A,z_A) = \frac{\hbar}{\epsilon_o c^2} \int_0^{\infty} \frac{\rm d \xi}{2 \pi} \int_{-\infty}^{\infty} \frac{{\rm d} k_y}{2 \pi}  \int_{-\pi /d}^{\pi /d} \frac{{\rm d} k_x^{(0)}}{2 \pi} \nonumber \\
&& \times \sum_{j,j'}  \frac{\xi^2} {2 \kappa_{j'}} \alpha (i \xi)  \; e^ {2 \pi i (j-j') x_A/d}  \;  e^{-(\kappa_j + \kappa_{j'}) z_A }  \label{final} \\
&& \times \sum_{p, p'} \langle j, p | \mathcal R_S (k_x^{(0)}, k_y, i\xi) | j', p' \rangle \; \hat {\boldsymbol {\epsilon}}_ {p}^+  \cdot \hat {\boldsymbol {\epsilon}}_ {p'}^-{}' . \nonumber
\end{eqnarray}
The sums over the Brillouin zones $j$ and $j'$ run from $-\infty$ to $\infty,$  the polarizations $p,p'$ are either E or H 
and $\kappa_j=\sqrt{k_x^{\scriptstyle(j)}{}^2+k_y^2+\xi^2/c^2}$ [see Eq.~(\ref{kxj})]. 
The scalar products are calculated from the expression for the unit vectors given by Eqs. (\ref{epsilonH}) and  (\ref{epsilonE}). When replacing 
 $\omega\rightarrow i\xi,$ $\sqrt{\omega^2/c^2-k^2}\rightarrow i\kappa_j,$ we find
\begin{eqnarray}
\nonumber
\hat {\boldsymbol {\epsilon}}_ {\rm H}^+  \cdot \hat {\boldsymbol {\epsilon}}_ {\rm H}^-{}'&=&-\frac{k_x^{\scriptstyle(j)} k_x^{\scriptstyle(j')}+\kappa_j\kappa_{j'}}{\xi^2/c^2+k_y^2},\\
\nonumber
\hat {\boldsymbol {\epsilon}}_ {\rm E}^+  \cdot \hat {\boldsymbol {\epsilon}}_ {\rm E}^-{}'&=&1+ \frac{c^2k_y^2}{\xi^2}\left(1-\hat {\boldsymbol {\epsilon}}_ {\rm H}^+  \cdot \hat {\boldsymbol {\epsilon}}_ {\rm H}^-{}'\right),\\
\nonumber
\hat {\boldsymbol {\epsilon}}_ {\rm E}^+  \cdot \hat {\boldsymbol {\epsilon}}_ {\rm H}^-{}'&=&
-\hat {\boldsymbol {\epsilon}}_ {\rm H}^+  \cdot \hat {\boldsymbol {\epsilon}}_ {\rm E}^-{}'=
\frac{c^3k_y(k_x^{\scriptstyle(j')} \kappa_j+ k_x^{\scriptstyle(j)}\kappa_{j'})}{\xi(\xi^2+c^2k_y^2)}.
\end{eqnarray}

The potential  (\ref{final}) does not depend on $y_A$ and is a periodic function of $x_A$ (with the same period of the grating) as expected. 
By choosing the origin $x=0$ at the mid-point of one of the grooves, the matrix elements  $ \langle j', p' | \mathcal R_S  | j, p \rangle$
turn out to be real and consistent with the even parity of the surface profile. 
They must also satisfy reciprocity relations \cite{Carminati98} which read, for our polarization basis, 
$\kappa_j \langle j, p | \mathcal R_S (k_x^{(0)}, k_y, i\xi) | j', p' \rangle = (2\delta_{pp'}-1) \kappa_{j'} \langle - j', p' | \mathcal R_S (-k_x^{(0)}, -k_y, i\xi) | - j, p \rangle.$
By combining these properties, we may cast (\ref{final}) as $U(x_A,z_A)=\sum_{j,j'} C_{j,j'}(z_A) \,e^ {2 \pi i (j-j') x_A/d}$ in terms of real coefficients satisfying
  $C_{j,j'}(z_A) = C_{j',j}(z_A).$ Thus, Eq.~(\ref{final}) yields a real potential  
  satisfying the required relations $U(x_A,z_A)=U(-x_A,z_A)=U(d-x_A,z_A)$ (reflection symmetries with respect to the
  groove and plateau 
 mid-points). 
  
\section{Numerical results}

In this section, we present numerical results for the Casimir-Polder potential given by (\ref{final}). We consider a  Rb atom, whose 
dynamic polarizability  $\alpha (i \xi)$ is obtained from Ref.~\cite{Derevianko10},
 interacting with an intrinsic silicon grating.  The corresponding dielectric constant $\epsilon(i\xi),$ required for the evaluation of the grating reflection operator ${\cal R}_S,$ 
is computed from available data at real frequencies \cite{Palik98} by using a Kramers-Kronig relation \cite{Lambrecht00}. 

When evaluating  (\ref{final}), we need to truncate the sum over the Brillouin zones $j,j'$ at some finite value 
  $N_{\rm max},$ so that the number of zones is 
  $2N_{\rm max}+1.$ 
 The value of $N_{\rm max}$ required for a given accuracy decreases with increasing distance $z_A,$ because the factor $ e^{-(\kappa_j + \kappa_{j'}) z_A } $ 
 kills the contribution of large values of $j$ and $j'.$ By comparing  results obtained from different values of $N_{\rm max},$ we found that $N_{\rm max}=3$
was sufficient 
  to achieve an accuracy at the level of a percent for the 
 numerical examples considered below.

 It is instructive to compare our exact results with those obtained within the proximity force approximation (PFA) \cite{Deriagin68}, which here corresponds to computing the potential $U$ 
for a given surface profile from the potential $U^{(0)}$ for a planar surface 
 taken at the local atom-surface distance:  $U(x_A,y_A,z_A)\approx U^{(0)}[z_A-h(x_A,y_A)].$ 
 For the rectangular grating considered here, 
 deviations from PFA are quantified by the ratio
 \(
 \rho= U(x_A,z_A)/U^{(0)}(z_A)
 \)
when the atom is on top of a groove and by
\(
 \rho= U(x_A,z_A)/U^{(0)}(z_A-a)
 \)
when the atom is on top of a plateau. 

\begin{figure}[h]
\begin{center}
\includegraphics[scale=.5]{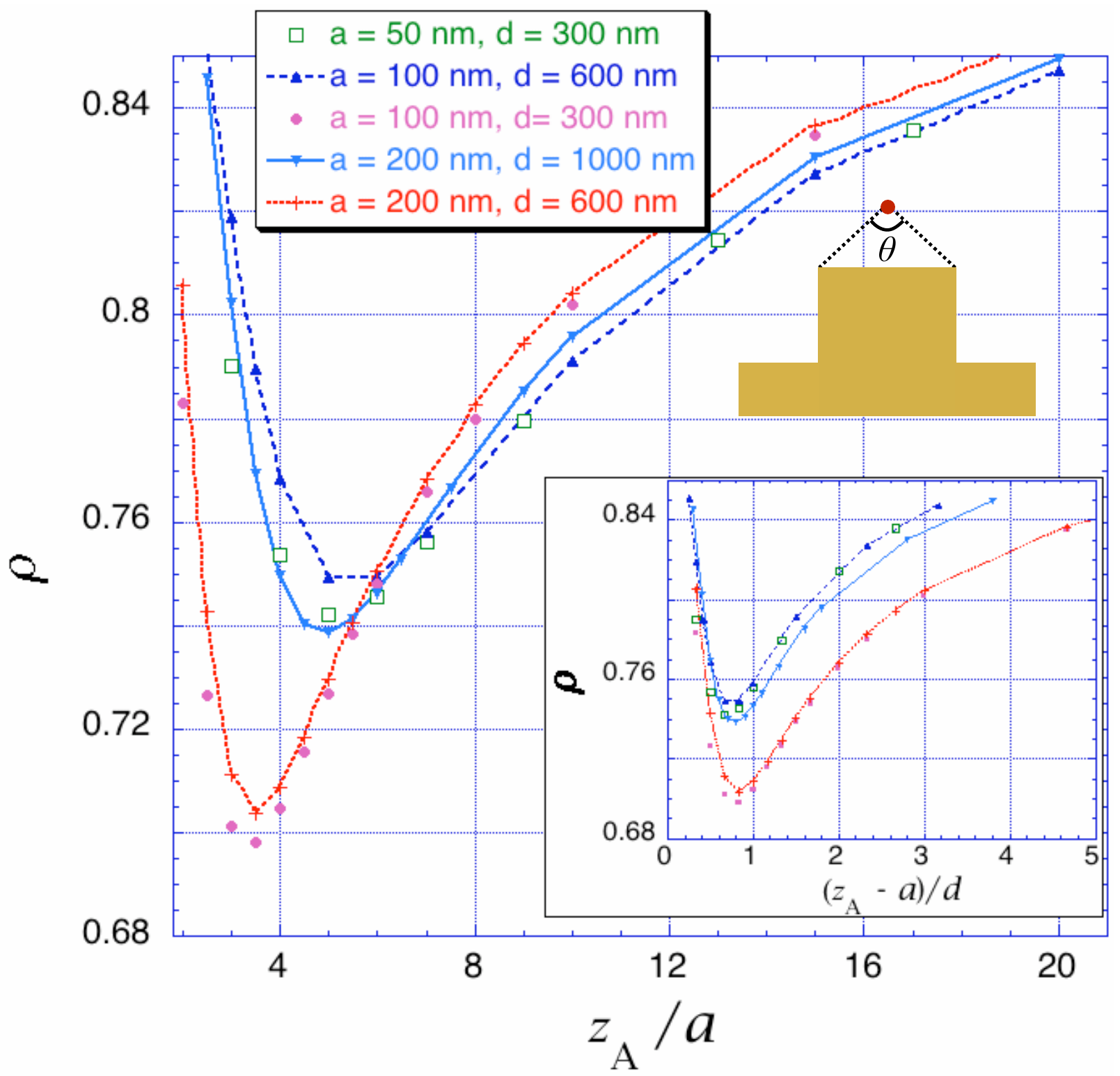} 
\end{center}
\caption{\label{plateau} \footnotesize{
Variation of $\rho=U(x_A,z_A)/U^{(0)}(z_A-a)$ versus $z_A/a$ and $(z_A-a)/d$ 
(inset)  for an atom above the plateau midpoint $x_A=d/2.$ We take fixed values of $a$ and $d$ for each set of data 
points. The angular aperture $\theta= 2\arctan[d/4(z_A-a)]$ shown in the diagram controls the departure from the PFA limit. 
}} 
\end{figure}

In Fig.~\ref{plateau}, we plot $\rho$ as a function of distance
 for  an atom above the plateau midpoint $x_A=d/2$ and
  several different combinations of amplitude and period, with groove width $s=d/2.$ We always find $\rho<1,$  in qualitative agreement 
with PWS,  with the grooves making
  the potential less attractive than the potential of a homogeneous plane surface at $z_A=a.$  
  When the atom is very far from the surface, $z_A\gg a,$ it feels the corrugation as a very small perturbation of the plane symmetry
   and then the potential approaches the value for a
  planar surface ($\rho=1$). As the distance decreases, the departure from the plane geometry becomes increasingly important.
 
The curves 
   corresponding to the same ratio $a/d$ are close but do not coincide, as expected since non-geometric length scales associated to
  characteristic frequencies of the Rb atom and the Si bulk 
   are also relevant. 
   For large distances, the dominant field frequencies are much smaller than the atom and medium characteristic frequencies, and then 
   the potential is dominated by the instantaneous response 
   associated to the  zero-frequency polarizability and dielectric constant. Thus, 
   the potential depends only on the ratios between the geometric lengths $a,$ $d$ and $z_A$ in this asymptotic limit.
   On the other hand, as the distance decreases, finite response times of both atom and medium give rise to a richer scenario. 
   The crossover between
    the long-distance Casimir-Polder and
    the unretarded van der Waals   
  regimes is in the range $200-300\,{\rm nm}$ for a Rb atom interacting with a Si surface (see for instance 
  Fig.~4 of Ref.~\cite{Messina09}). When the separation distance decreases below  $z_A/a=5,$ it already corresponds to the transition to the van der Waals regime for
  $a=50\,{\rm nm}$ but not yet for $a=100 \,{\rm nm}.$ 
  The values for $a=50\,{\rm nm}$ then move away from the curve for 
  $a=100 \,{\rm nm}$ and the same 
   $a/d=1/6$ as shown in Fig.~\ref{plateau}. A similar effect, for $a=100 \,{\rm nm}, 200 \,{\rm nm}$  
   and $a/d=1/3$ is also apparent in Fig.~\ref{plateau}.

  As $\rho$ approaches the PFA limit $\rho\rightarrow 1$ at shorter distances, the amplitude $a$ is no longer the most relevant length scale capturing the 
  variation with distance.  
  In the inset  of Fig.~\ref{plateau}, we plot
   $\rho$ as a function
   of   $(z_A-a)/d,$ which is 
   directly related to the 
   angular aperture $\theta= 2\arctan[d/4(z_A-a)]$ of the plateau width as seen from the atom location above the plateau midpoint
  (see diagram in 
   Fig.~\ref{plateau}). When $\theta$ is very close to $180^{\rm o}$ [$(z_A-a)/d\ll 1$], 
   we expect the effect of the border of the plateau  to be 
   negligible, and then 
   the potential should be well approximated by the result for an infinite plane (PFA). 
      The inset of Fig.~\ref{plateau} shows that the variable  $(z_A-a)/d$ indeed captures  the main effect behind the departure from PFA as the 
      atom is displaced away from the surface, since 
      the curves corresponding to different values of  amplitude and 
   period collapse near each other at short distances.
      For all different values of  $a$ and $d$ shown in the figure, the point of maximum  $1-\rho$ is near $(z_A-a)/d\sim 0.8,$ corresponding to an angular aperture 
   $\theta \approx 35^o.$
  The maximum  $1-\rho$ depends mainly on $a/d$ and is larger for larger $a/d$ as expected.

\begin{figure}[h]
\begin{center}
\includegraphics[scale=.45]{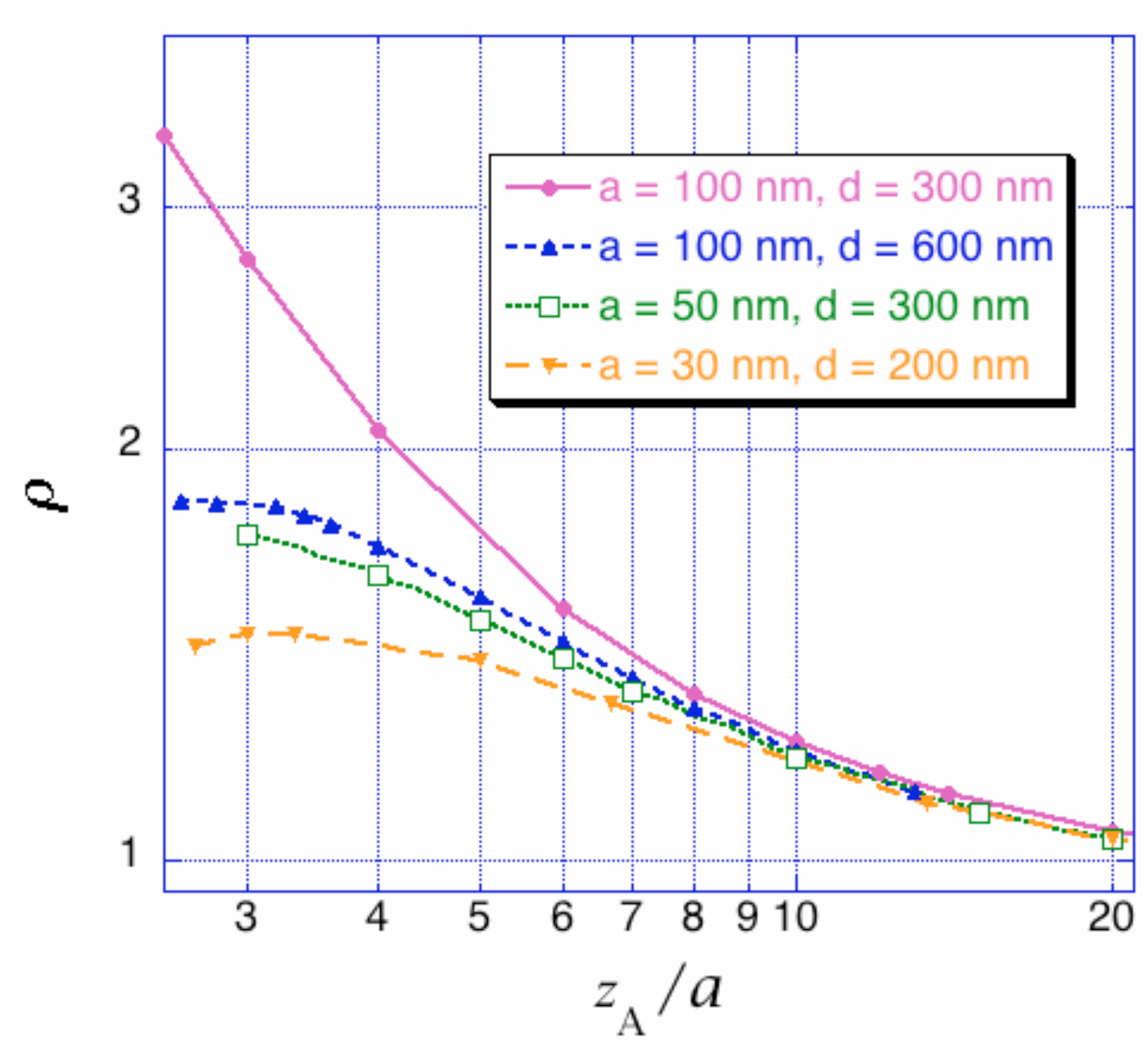} 
\end{center}
\caption{\label{groove} \footnotesize{Variation of $\rho=U(x_A,z_A)/U^{(0)}(z_A)$ versus $z_A/a$ for an atom above the groove midpoint $x_A=0.$}} 
\end{figure}

When the atom is above the groove region, the potential is stronger than the potential for a planar surface at the bottom of the groove ($\rho>1$), 
again in qualitative agreement with the PWS picture. In Fig.~\ref{groove}, we plot $\rho$ as a function of distance for $x=0$ (groove midpoint) and 
$s=d/2.$
The curves corresponding to different (fixed) values of 
$a$ and $d$  
merge at distances $z_A/a\stackrel{>}{\scriptscriptstyle\sim} 8.$ 
This remarkable property was first found in Ref.~\cite{Dobrich08} in the context of the  scalar Casimir-Polder  model. It may be interpreted 
as an effect of 
averaging the small length scales  associated to the surface profile when considering the long-wavelength field fluctuations that provide the 
main contribution at large distances. 

At shorter distances, the curves move apart following the order of larger $a/d.$ Although 
atomic/material frequency scales play a relevant
 role at $z_A< 300\,{\rm nm},$ the values for different $a$ and $d$
with the same $a/d$ shown in Fig.~\ref{groove} are still relatively close.

\begin{figure}[h]
\begin{center}
\includegraphics[scale=.35]{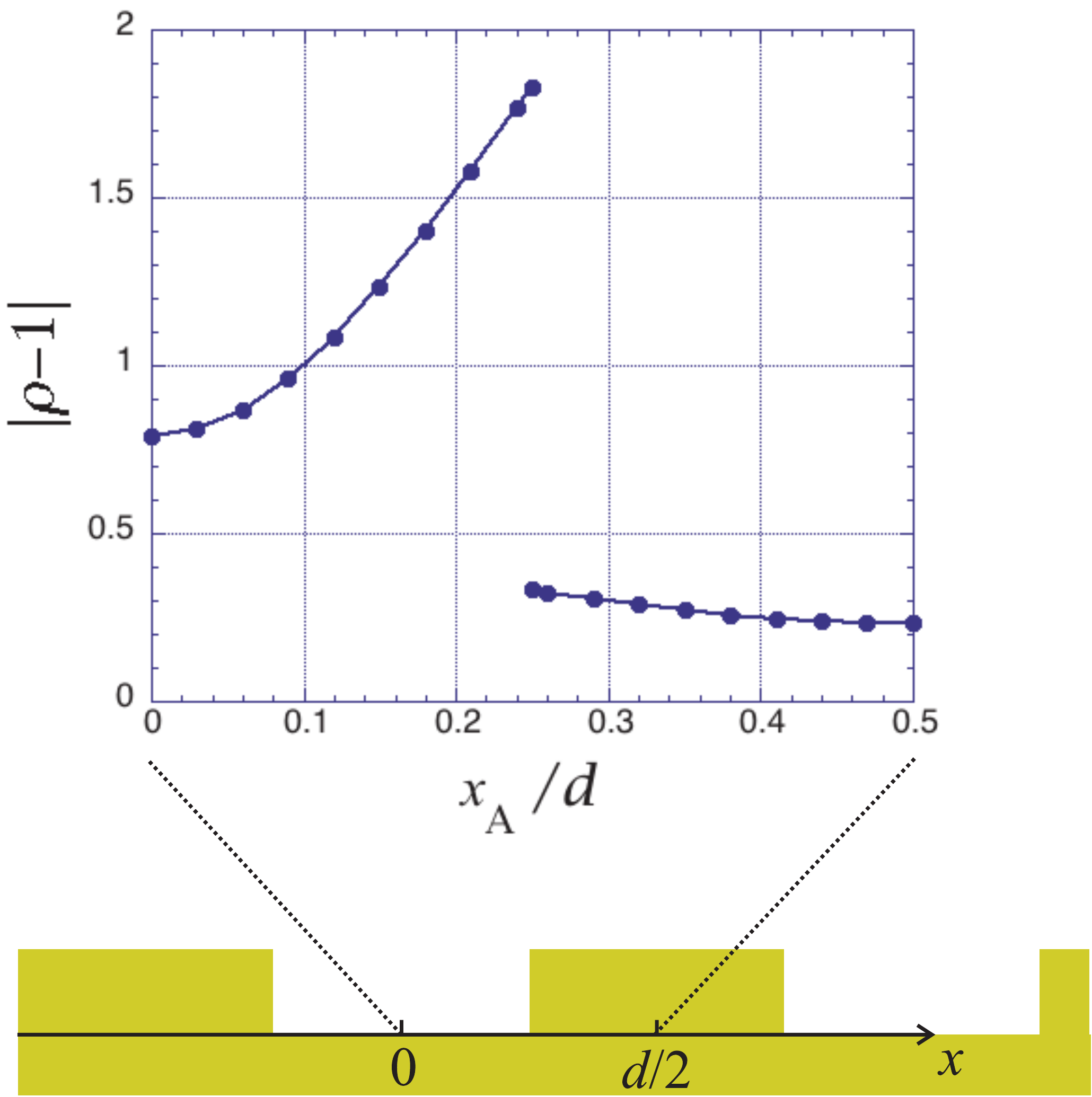} 
\end{center}
\caption{\label{rho_vs_x} \footnotesize{Lateral variation of  $|\rho-1|$ at a fixed local distance $z_A-h(x_A)$
with $a=100\,{\rm nm}$ and $d=600\,{\rm nm}.$ 
We take $z_A= 3a$
above the groove ($0\le x_A<d/2$),  and  $z_A= 4a$ above the plateau
($d/2\le x_A<d$).
}} 
\end{figure}

From an overall comparison of 
 Figs.~\ref{plateau} and \ref{groove}, we may conclude that 
the relative deviation from PFA $|\rho-1|$ is generally larger above the groove than above
 the plateau. In Fig.~\ref{rho_vs_x}, we plot $|\rho-1|$ as a function of the lateral position 
 for a fixed local distance $z_A-h(x_A)=3a,$ with $a=100\,{\rm nm},$ $d=600 \,{\rm nm},$
and $s=300 \,{\rm nm}.$ PFA is indeed worse above the groove, where the deviation increases rapidly as one approaches the corner. 
Above the plateau, PFA is also worse near the corner, but the variation of $\rho$ with lateral position is much smaller.

\begin{figure}[h]
\begin{center}
\includegraphics[scale=.55]{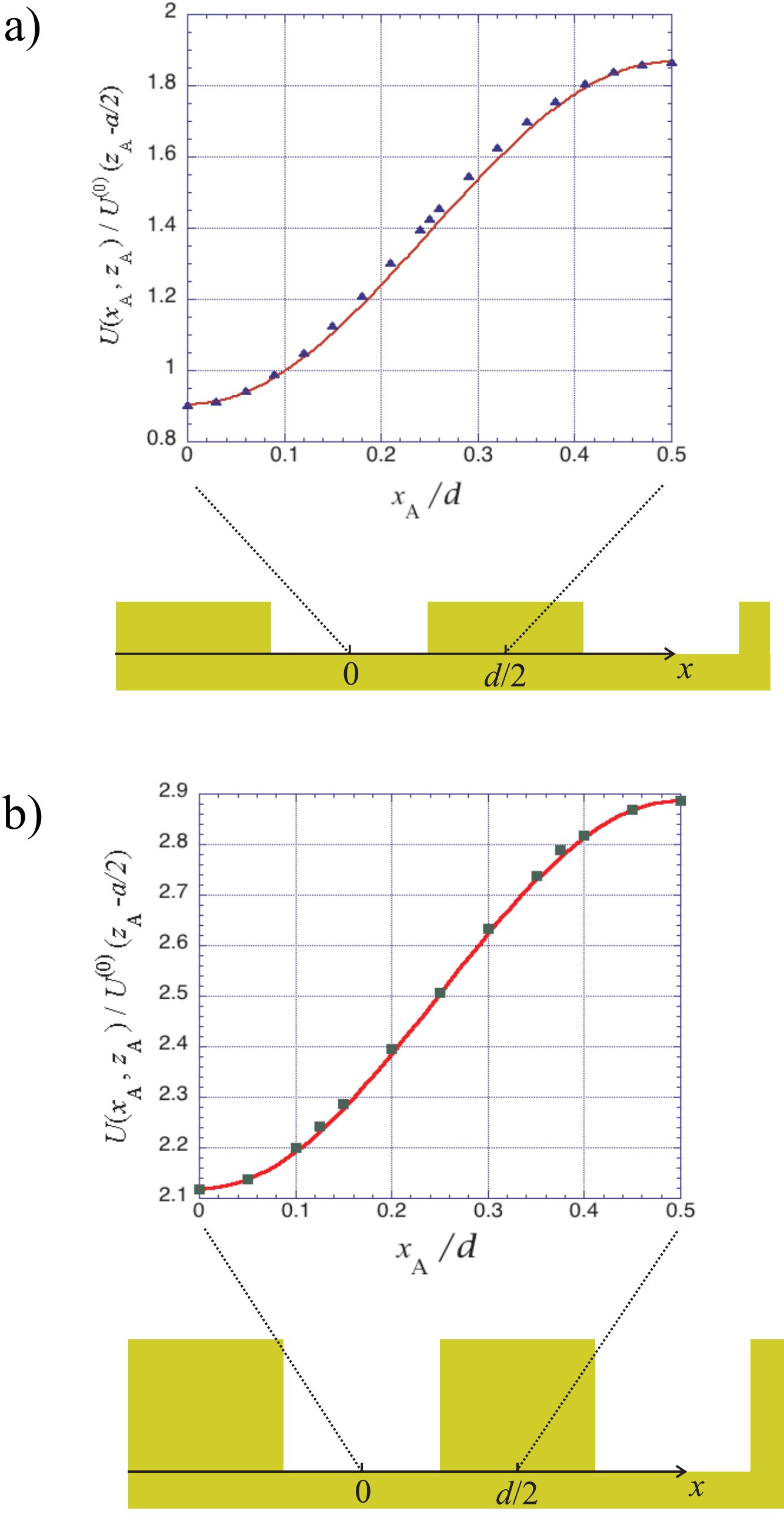} 
\end{center}
\caption{\label{variando_x} \footnotesize{Lateral variation of the Casimir-Polder potential
divided by $U^{(0)}(z_A-a/2)$
 for  $a=100\,{\rm nm}$ with  (a)  $d=2z_A=6 a;$ and
(b) $d=z_A=2a.$ The solid lines represent the sinusoidal function of  period $d$ 
that coincides with the exact potential at the groove and plateau midpoints.}} 
\end{figure}

At a given plane $z=z_A,$ the potential profile along the $x$ axis is approximately
 sinusoidal for the relatively large distances considered here.  In   Fig.~\ref{variando_x}a, we compare 
 the values for $U(x_A,z_A)/U^{(0)}(z_A-a/2)$ with a sinusoidal variation (solid line) corresponding to period $d$ and 
 amplitude $[U(d/2,z_A)-U(0,z_A)]/2.$ We use $z_A=3a$ and the same parameters employed in Fig.~\ref{rho_vs_x}. 

At very large distances, the main contribution comes from specular reflection at 
the first Brillouin zone $j=j'=0,$ yielding a flat potential according to Eq.~(\ref{final}) (see  Figs. \ref{plateau} and \ref{groove}).
As the distance decreases, the major correction, corresponding to the first diffraction order $j-j'=\pm 1,$
produces a sinusoidal modulation with period $d.$
 Deviation from the sinusoidal shape (coming from $|j-j'|\ge 2$) is such that
the lateral force stiffness $|\partial^2U/\partial x_A^2|$
 is slightly larger at the groove midpoint than at the plateau midpoint, as shown in Fig.~\ref{variando_x}a.

The deviation from the sinusoidal shape and the amplitude of oscillation are controlled by the angular aperture parameter $(z_A-a)/d,$ 
whereas $z_A/a$ controls the convergence of the spatial average  $\langle U(x_A,z_A) \rangle$   to 
its large distance limit  $U^{(0)}(z_A-a/2) = [U^{(0)}(z_A-a)+U^{(0)}(z_A)]/2 +{\cal O}(a/z_A)^2$ when $s=d/2.$
For instance, in Fig.~\ref{variando_x}b we plot the potential as a function of $x_A/d$ for $z_A=d=2 a= 200\,{\rm nm}.$ 
These parameters correspond to $z_A/a=2$ and $(z_A-a)/d=0.5,$ to be compared with $z_A/a=3$ and $(z_A-a)/d= 0.33$ for Fig.~\ref{variando_x}a. 
While the average potential is closer to the asymptotic value 
$U^{(0)}(z_A-a/2)$ in Fig.~\ref{variando_x}a (larger $z_A/a$), the amplitude is smaller and the shape closer to the sinusoidal fit 
 in Fig.~\ref{variando_x}b (larger  $(z_A-a)/d$).  As the atom is displaced away from the surface, 
  the potential becomes more
sinusoidal  and with a smaller  
 amplitude. With the same parameters of Fig.~\ref{variando_x}b,  the amplitude decreases
  by two and three  orders of magnitude at $z_A=3 a$ and $z_A=4 a,$ respectively.

\section{Concluding remarks}

We have shown that 
PFA overestimates the potential strength above the plateau. As the atom is displaced away from surface, the deviation  $1-\rho$ is mainly controlled by 
the  parameter $(z_A-a)/d$ associated to the angular aperture of the plateau as seen by the atom. 
 The maximum deviation takes place near $(z_A-a)/d=0.8$ and is larger for deeper grooves (larger $a/d$) as expected.
As the distance is increased past this value, the potential again approaches the result for a planar surface.  
On the other hand, the potential is underestimated by the PFA above the groove. The deviation increases strongly 
as the atom is laterally displaced closer to a corner,  where the presence of the  ridge walls have a 
stronger effect. 

The naive PWS picture of the Casimir-Polder interaction  agrees qualitatively with several features found in this paper. Nevertheless, 
PWS typically underestimates 
 the magnitude of the Casimir-Polder potential in the grating geometry  \cite{Dalvit08},
 particularly in the  non-perturbative regime considered here \cite{Impens10}. For instance, it predicts an anisotropy of the Casimir-Polder potential far too small
 to nucleate vortices in a Bose-Einstein condensate by rotation of the grating, whereas the exact theory presented here predicts an anisotropy above the required values 
 under realistic experimental conditions~\cite{Impens10}. 

We have shown that the contribution of higher diffraction orders to the potential is increasingly small as  the distance increases, 
resulting in a sinusoidal variation along the transverse $x$ direction for  
$(z_A-a)/d\stackrel{>}{\scriptstyle\sim} 0.5$ (aperture angle smaller than $53^o$).  

The scattering approach developed in this paper might be readily adapted to consider more general two-dimensional periodic patterns, like the pillars 
structure employed in the quantum reflection experiment reported by Ref.~\cite{Pasquini06}. 
Recently, the Casimir interaction between two  material surfaces imprinted with two-dimensional periodic patterns has been analyzed in detail~\cite{Davids10}.  
The method developed here thus paves the way for 
 the quantitative analysis of a variety of interesting geometries with applications in  enhanced atomic quantum reflection and diffraction. 

\begin{acknowledgments}

We would like to thank Fran\c cois Impens and Valery Marachevsky for  discussions.
This work was partially supported by CAPES-COFECUB, CNPq, DARPA, 
ESF Research Networking Programme CASIMIR and  FAPERJ-CNE.

\end{acknowledgments}

\end{document}